\documentclass[aps,pre,twocolumn,showkeys,showpacs]{revtex4-1}
\usepackage[OT4]{fontenc}
\usepackage[colorlinks=true,hyperfootnotes=true,breaklinks=true]{hyperref}
\usepackage{auto-pst-pdf}
\usepackage{graphicx} 
\usepackage{psfrag}

\begin{document}

\title{Influence of long-range interactions on strategy selection in crowd}\thanks{Presented at the \href{http://summersolstice2013.if.pw.edu.pl/}{Summer Solstice 2013 International Conference on Discrete Models of Complex Systems}, Warsaw, Poland, Jun. 27-29, 2013.}

\author{Krzysztof Malarz}  
\homepage{http://home.agh.edu.pl/malarz/} 
\email{malarz@agh.edu.pl} 
\author{Ma{\l}gorzata J. Krawczyk}  
\author{Krzysztof Ku{\l}akowski}
\affiliation{\href{http://www.agh.edu.pl/}{AGH University of Science and Technology},
\href{http://www.pacs.agh.edu.pl/}{Faculty of Physics and Applied Computer Science},
al. Mickiewicza 30, 30-059 Krakow, Poland.}

\date{\today}

\begin{abstract}
An order--disorder phase transition is observed for Ising-like systems even for arbitrarily chosen probabilities of spins flips [K. Malarz et al, Int. J. Mod. Phys. 
C {\bf 22}, 719 (2011)]. For such athermal dynamics one must define $(z+1)$ spin flips probabilities $w(n)$, where $z$ is a number of the nearest-neighbours for 
given regular lattice and $n=0,\cdots,z$ indicates the number of nearest spins with the same value as the considered spin. Recently, such dynamics has been successfully 
applied for the simulation of a cooperative and competitive strategy selection by pedestrians in crowd [P.~Gawro\'nski et al, Acta Phys. Pol. A {\bf 123}, 522 (2013)]. 
For the triangular lattice ($z=6$) and flips probabilities dependence on a single control parameter $x$ chosen as $w(0)=1$, $w(1)=3x$, $w(2)=2x$, $w(3)=x$, $w(4)=x/2$, 
$w(5)=x/4$, $w(6)=x/6$ the ordered phase (where most of pedestrians adopt the same strategy) vanishes for $x>x_C\approx 0.429$. In order to introduce long-range interactions 
between pedestrians the bonds of triangular lattice are randomly rewired with the probability $p$. The amount of rewired bonds can be interpreted as the probability of 
communicating by mobile phones. The critical value of control parameter $x_C$ increases monotonically with the number of rewired links $M=pzN/2$ from $x_C(p=0)\approx 0.429$ 
to $x_C(p=1)\approx 0.81$.
\end{abstract}



\maketitle

\section{Introduction}

In theoretical studies the critical point  $x_C$ (i.e. Curie temperature $T_C$ for Ising model, or percolation threshold $p_C$ in geometrical systems) may by influenced by 
lattice/network topology \cite{archi,archi-1,archi-2,archi-3,archi-4}, numerical scheme of spin updates \cite{sousa}, clustering coefficient of the network \cite{manka}, range of 
interaction \cite{solomon} and also by assumed sites neighbourhood for geometrical systems \cite{percol,percol-1,percol-2,percol-3}. Here we consider an order-disorder transition 
in an athermal system, where the probabilities of change of a local (spin-like) variable depends in arbitrary way on the system parameters. In this
system, the concepts of energy and temperature do not apply. Recently, an order--disorder phase transition has been observed for such a system \cite{korff}. For such 
athermal dynamics one has to define $(z+1)$ spin flips probabilities $w(n)$, where $z$ is a number of the nearest-neighbours for given regular lattice and $n=0,\cdots,z$ 
indicates the number of nearest spins with the same value as the considered spin. This dynamics has been successfully applied for the simulation of a cooperative and competitive 
strategy selection by pedestrians in crowd \cite{fens6}. For the triangular lattice ($z=6$) and flips probabilities dependence on a single control parameter $x$ chosen as $w(0)=1$, $w(1)=3x$, $w(2)=2x$, $w(3)=x$, $w(4)=x/2$, $w(5)=x/4$, $w(6)=x/6$ the ordered phase (where most of pedestrians adopt the same strategy) vanishes for $x>x_C\approx 0.429$.

\begin{figure}
\begin{center}
\includegraphics[width=0.47\textwidth]{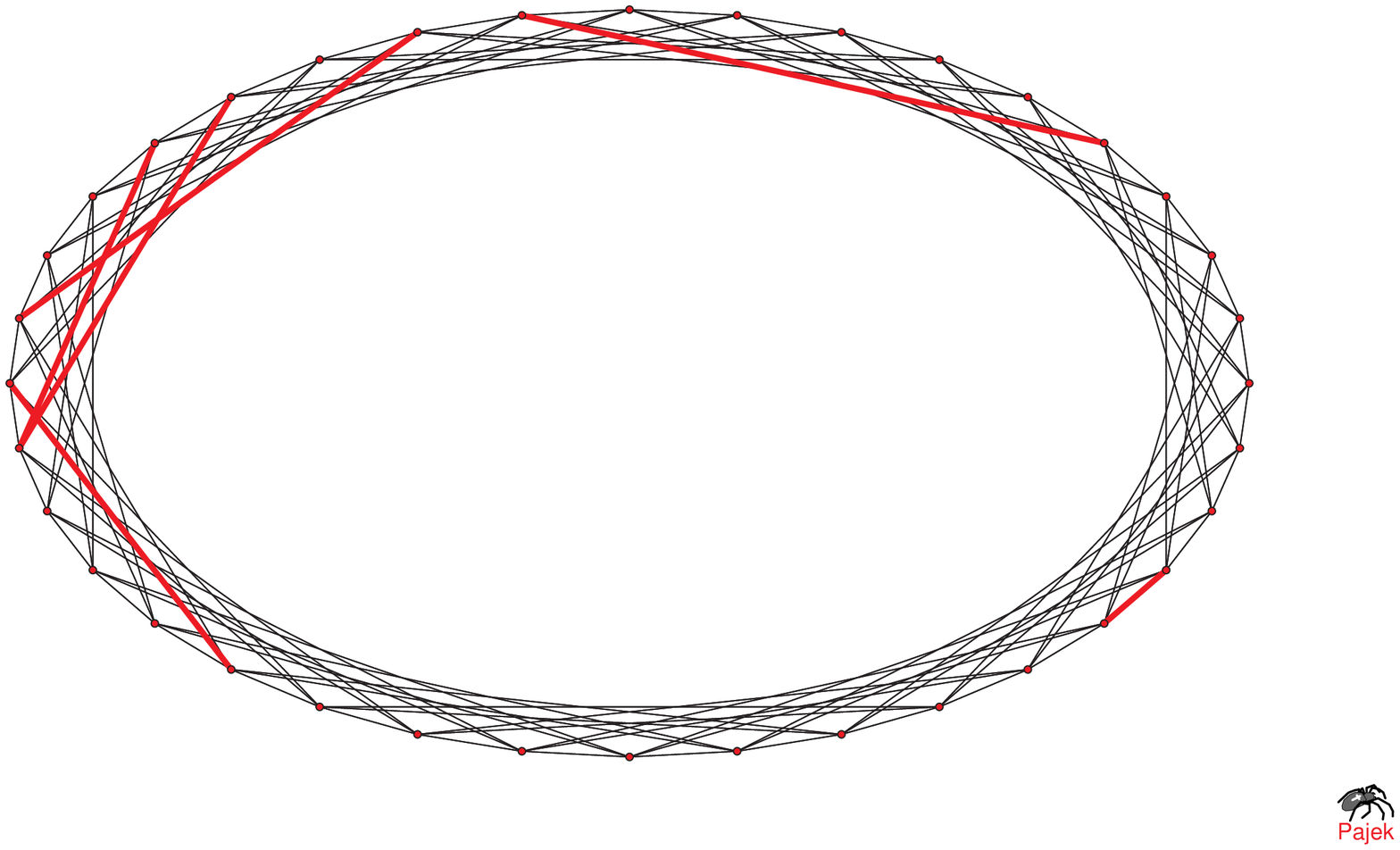}
\includegraphics[width=0.47\textwidth]{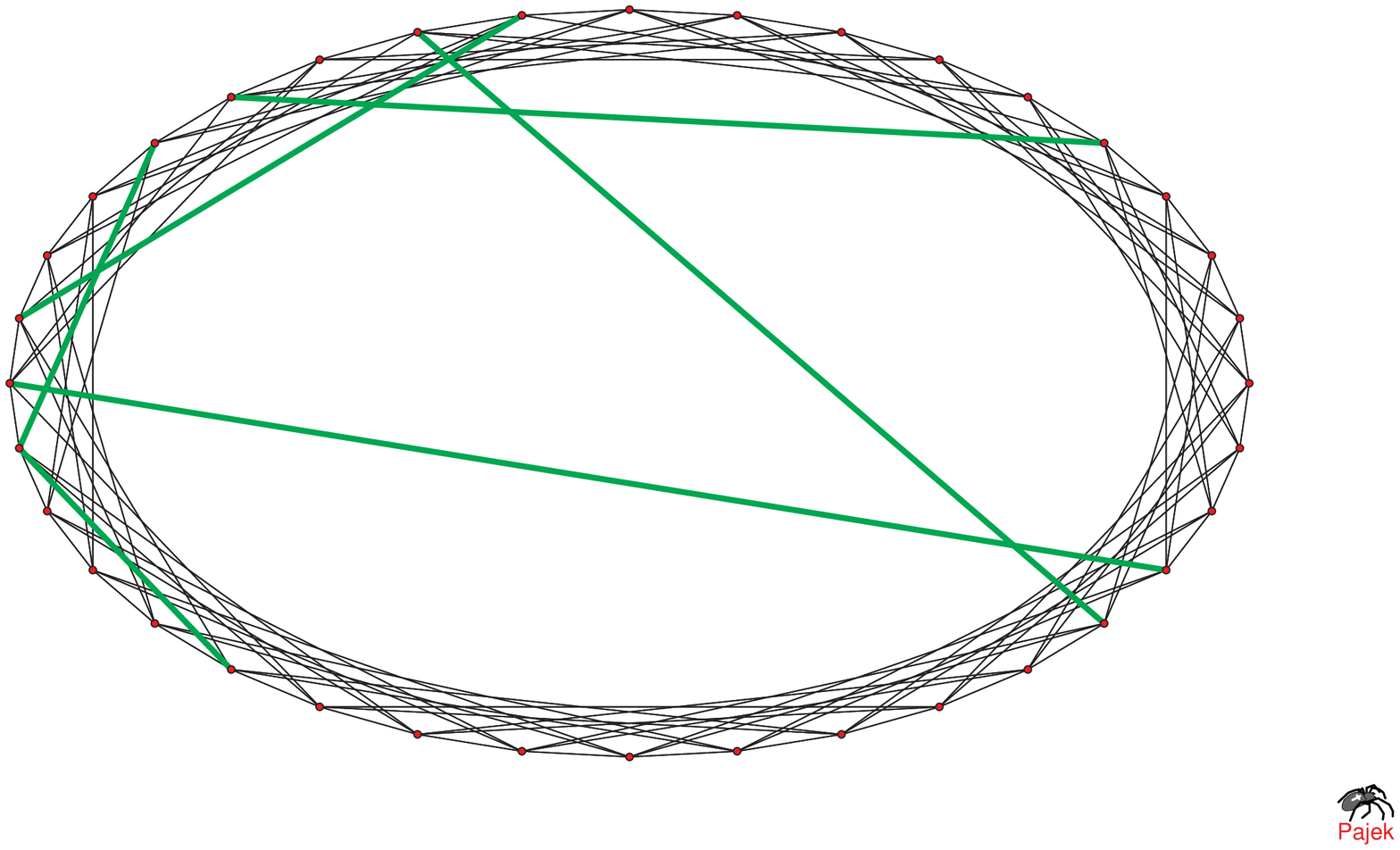}
\end{center}
\caption{\label{net} Sketch of network construction. 
The red links are removed from triangular lattice (here with helical boundary conditions) and replaced by green ones with rewiring probability $p$.
Graph was prepared with Pajek software \cite{pajek}.} 
\end{figure}

In this paper we extend our recent studies \cite{fens6} by introducing long-range interactions among pedestrians in a crowd.
In order to introduce long-range interactions between pedestrians the bonds of triangular lattice are randomly rewired with the probability $p$.
The schematic sketch of network construction is presented in Fig. \ref{net}.

We show that critical value of control parameter $x_C$ increases monotonically with the number of rewired links $M=pzN/2$ from $x_C(p=0)\approx 0.429$ to $x_C(p=1)\approx 0.81$.
Moreover, we present others signatures of order--disorder phase transition occurrence, including the Binder cumulant $U_4$ and pedestrians' susceptibility for changing their strategy $\chi$ behaviours in the vicinity of phase transition.

\section{Model}
The system contains $N$ sites of triangular lattice with helical boundary conditions (see Fig. \ref{net}).
Each lattice node is decorated with a single spin-like variable $s_i=\pm 1$ representing actual strategy (i.e. cooperative or competitive) adopted by a pedestrian $i$ in a crowd.
The long-range interactions among pedestrian are introduced by random rewiring of $M=pzN/2$ links, where $p$ is the single edge rewiring probability.
In every Monte Carlo step each pedestrian is investigated either he/she will change his/her strategy ($s_i(t+1) = -s_i(t)$) or not ($s_i(t+1) = s_i(t)$).
The probabilities of changing mind by pedestrians are given as $w(n)$, where $n$ indicates the number of the nearest pedestrian using the same strategy as the considered agent $i$.
We use the same set of probabilities as in Ref. \cite{fens6}, i.e.: $w(0)=1$, $w(1)=3x$, $w(2)=2x$, $w(3)=x$, $w(4)=x/2$, $w(5)=x/4$, $w(6)=x/6$, where $x$ is a model control parameter.

After reaching by the system an equilibrium state during the first $T$ Monte Carlo steps we compute temporal average of the order parameter and its higher moments
\[ \langle m^k\rangle = T^{-1} \sum_{t=T+1}^{2T} [m(t)]^k, \qquad k=1,2,4, \]
where
\[ m(t)=N^{-1}\sum_{i=1}^N s_i(t) \]
is a spatial average of the pedestrian strategies and $2T=10^6$, $10^6$, $10^7$, $10^7$, $10^8$, $10^8$ for $N=10^6$, $512^2$, $256^2$, $128^2$, $64^2$ and $32^2$, respectively.

To observe additional signatures of the order--disorder phase transition in our system we evaluate the fourth-order Binder cumulant
\begin{equation}
\label{eq_U4}
U_4=1-\frac{\langle m^4\rangle}{3\langle m^2\rangle ^2}
\end{equation}
and pedestrians susceptibility for changing opinion
\begin{equation}
\label{eq_chi}
\chi=\frac{dm}{dh}.
\end{equation}

In the latter definition the equivalent of external magnetic field $h$ could play a role of common agents believes that using one of the strategy (for instance cooperative) may be better than using other one (aggressive, competitive and selfish).
Thus gentlemen will not push other gentlemen, ladies and children just to have more comfortable way to the nearest exit.
On contrary a group of football hooligans may find previously described strategy as strange and useless. The probabilities $w(n)$ must be redefined as $w_\pm(n)=w(n)\mp h$ in order to introduce above mentioned effects.
Then $w_+$ and $w_-$ correspond to the probabilities for agents using $s_i=+1$ and $s_i=-1$ strategies, respectively.
After such modification and assuming $h>0$ all agents using $s_i=+1$ strategy will adopt opposite strategy with a lower probability, while agents using opposite strategy ($s_i=-1$) will change it more likely in contrast to situation with $h=0$.
If for some combination of $x$ and $h$ values of probabilities $w(n)$, $w_+(n)$ or $w_-(n)$ are greater than one (less than zero) then we assume that they are equal to one (zero).

\section{Results}

\begin{figure}
\begin{center}
\psfrag{m(t)}{$m(t)$}
\psfrag{t}{$t$}
\psfrag{x=}{$x=$}
\psfrag{N=10^6, p=1/2}{$p=1/2$}
\includegraphics[width=0.47\textwidth]{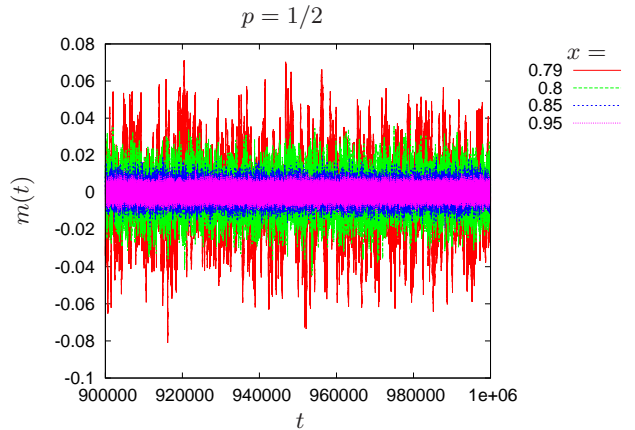}
\end{center}
\caption{\label{mvst}
Temporal dependence of order parameter $m(t)$ for various values of parameter $x>x_C$ and rewiring intensities $p$.
The latter do not influence the results qualitatively.
The simulations are carried out for lattice with $N=10^6$ sites. 
The last $10^5$ time steps are displayed.} 
\end{figure}

In the vicinity of the phase transition $x_C$ the critical slowing down was observed for original unrewired lattice ($p=0$) \cite{fens6}.
It means that when model control parameter $x$ approaches the critical point $x\to x_C^+$ the order parameter $m(t)$ oscillations become more intensive.
Introducing of long-range interactions does not destroy this effect as presented in Fig. \ref{mvst}.

\begin{figure}
\begin{center}
\psfrag{<m>}{$\langle m\rangle$}
\psfrag{<m^2>}{$\langle m^2\rangle$}
\psfrag{x}{$x$}
\psfrag{p=}{$p=$}
\includegraphics[width=0.47\textwidth]{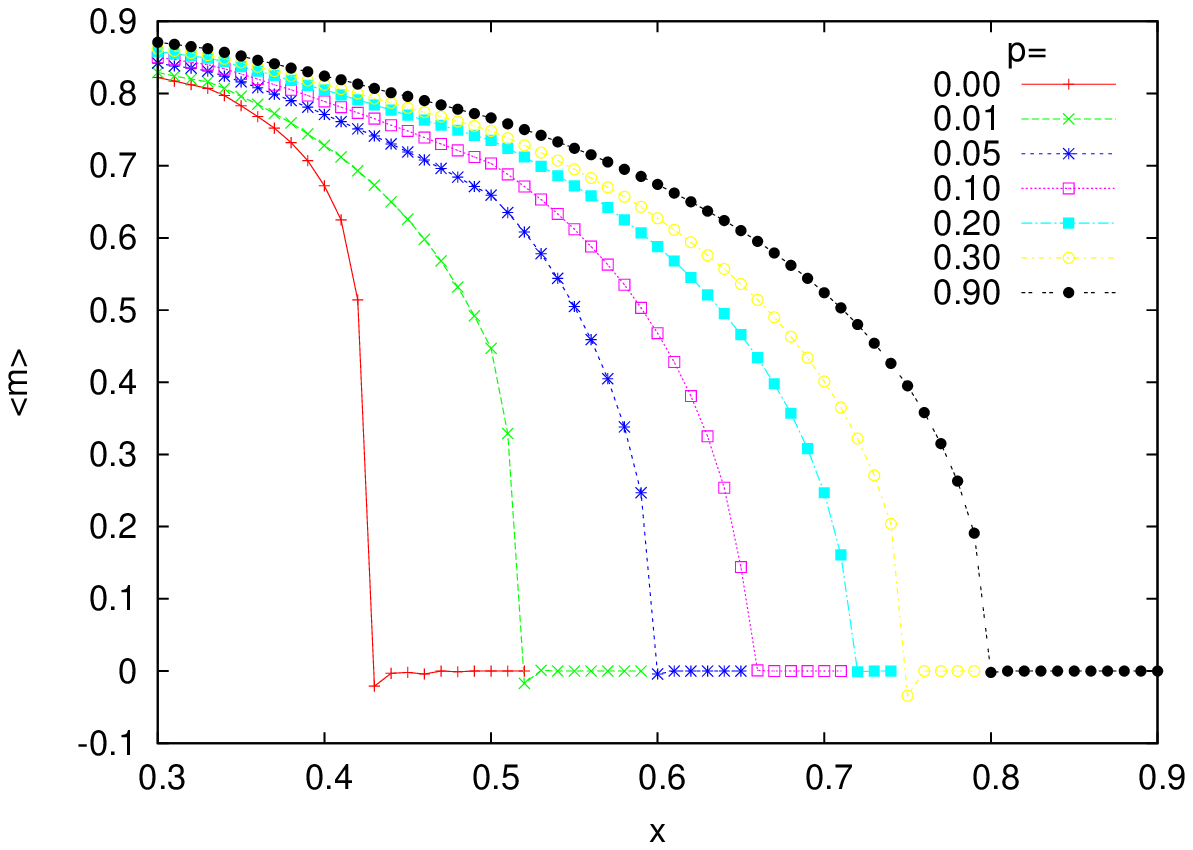}\\
\includegraphics[width=0.47\textwidth]{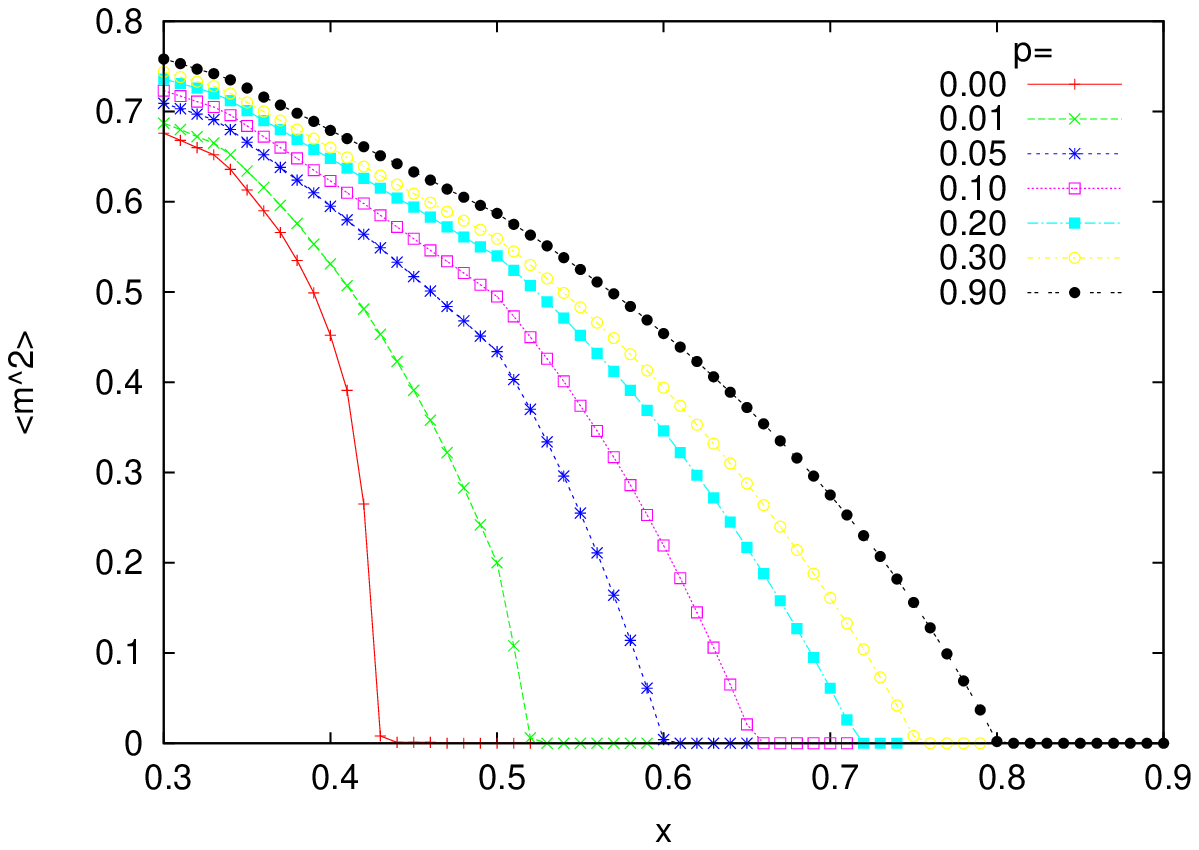}
\end{center}
\caption{\label{mvsx} Order parameter $\langle m\rangle$ and $\langle m^2\rangle$ dependence on model control parameter $x$ for various rewiring probabilities $p$.
The temporal average over last $T=5\times 10^5$ sweeps through the lattice is used to evaluate average values of $\langle m\rangle$ and $\langle m^2\rangle$.
The values of $x$ for which $\langle m^2\rangle$ decrease to zero approximate the critical values of $x_C$.
The simulations are carried out for lattice with $N=10^6$ sites.} 
\end{figure}

\begin{figure}
\begin{center}
\psfrag{x_c}{$x_C$}
\psfrag{p}{$p$}
\setlength\fboxsep{10pt}
\setlength\fboxrule{0.0pt}
\fbox{
\includegraphics[width=0.47\textwidth]{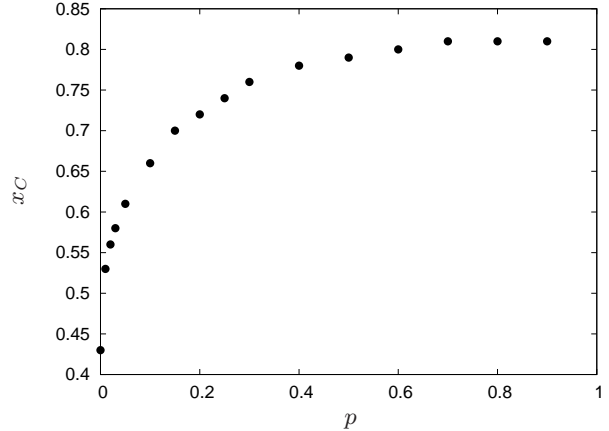}
}
\end{center}
\caption{\label{xcvsp} Critical value of the model control parameter $x_C$ dependence on a rewiring probability $p$.} 
\end{figure}

In Fig. \ref{mvsx} the temporal order parameter $\langle m\rangle$ and $\langle m^2\rangle$ dependence on model control parameter $x$ are presented.
The value of $x$ parameter for which $\langle m\rangle$ and $\langle m^2\rangle$ vanish correspond to the critical point $x_C$.
The dependence of critical value of the model control parameter $x_C$ on rewiring probability $p$ is presented in Fig. \ref{xcvsp}.
The critical value of control parameter $x_C$ increases monotonically with the number of rewired links $M=pzN/2$ from $x_C(p=0)\approx 0.429$ to $x_C(p=1)\approx 0.81$.

\begin{figure*}
\begin{center}
\psfrag{(a) p=0.01}{(a) $p=0.01$}
\psfrag{(b) p=0.9}{(b) $p=0.9$}
\psfrag{(c) p=0.01}{(c) $p=0.01$}
\psfrag{(d) p=0.9}{(d) $p=0.9$}
\psfrag{x}{$x$}
\psfrag{xC}{$x_C$}
\psfrag{chi}{$\chi$}
\psfrag{U_4}{$U_4$}
\psfrag{x}{$x$}
\psfrag{L=}{$\sqrt{N}=$}
\includegraphics[width=0.47\textwidth]{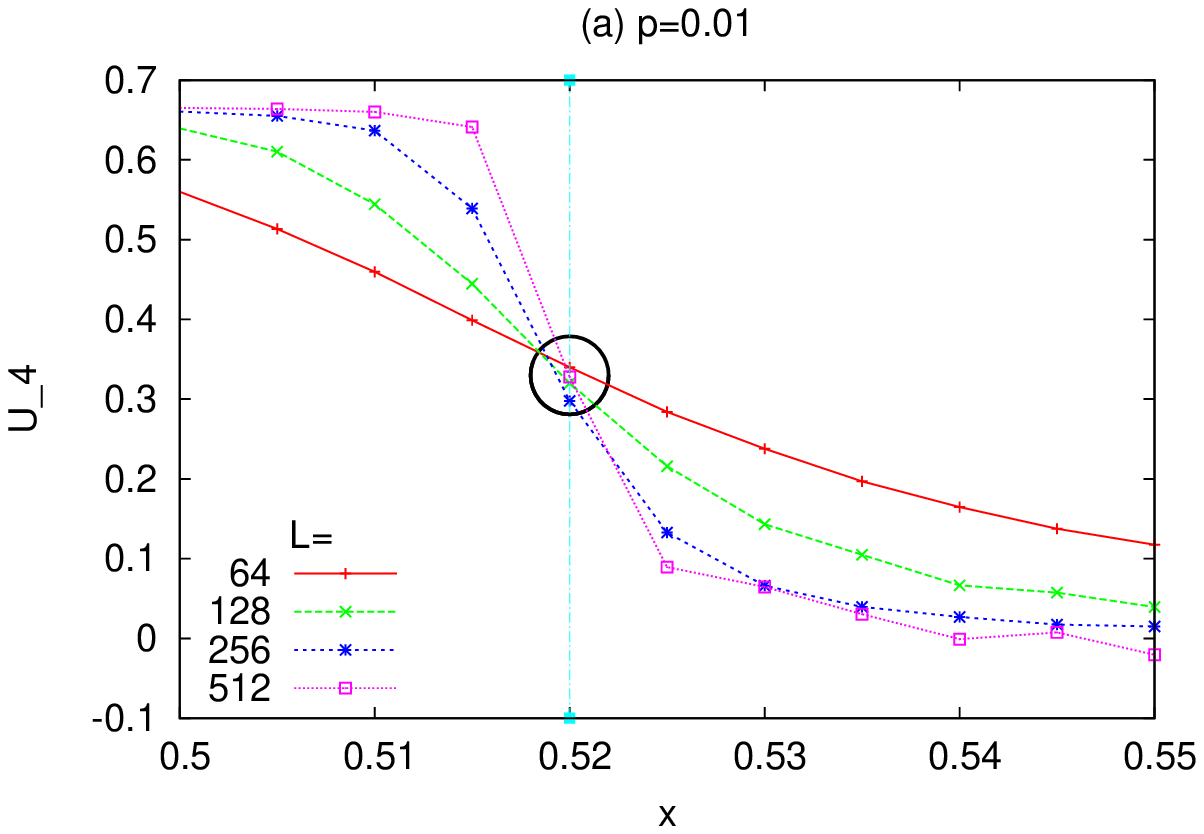}
\includegraphics[width=0.47\textwidth]{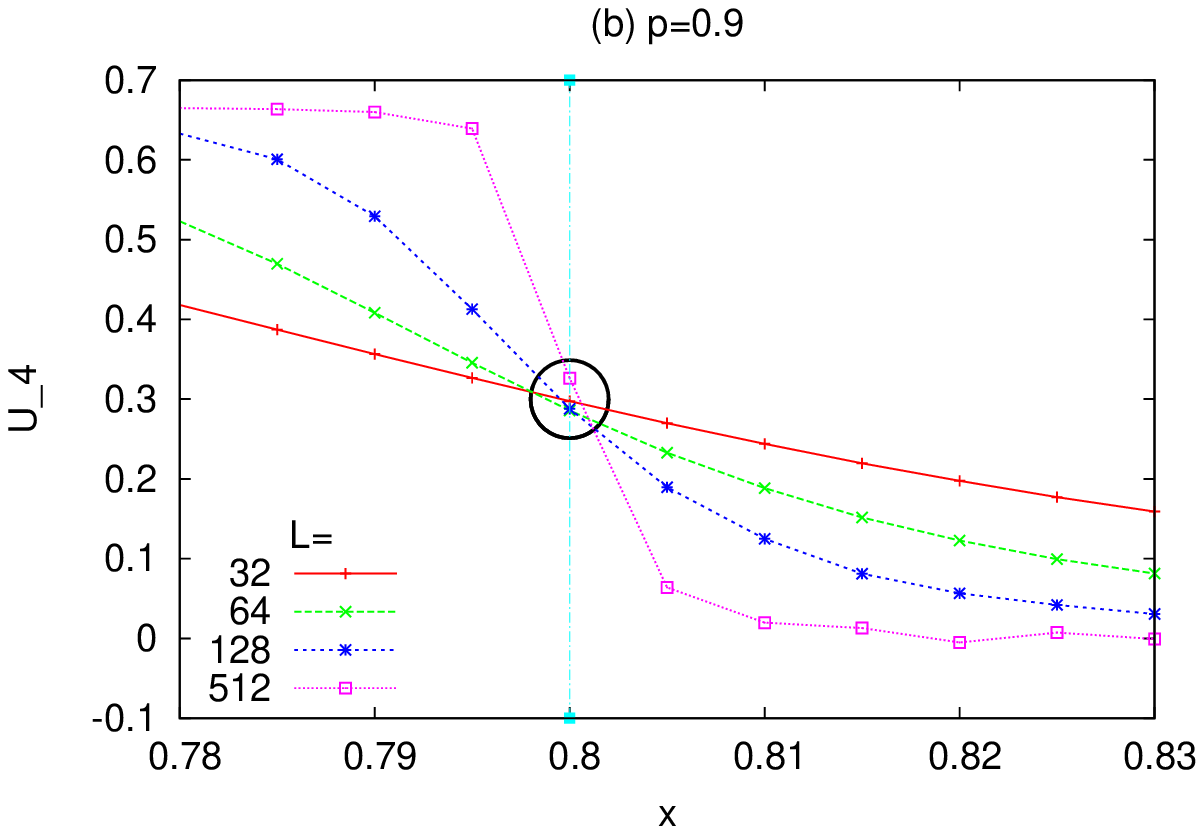}
\includegraphics[width=0.47\textwidth]{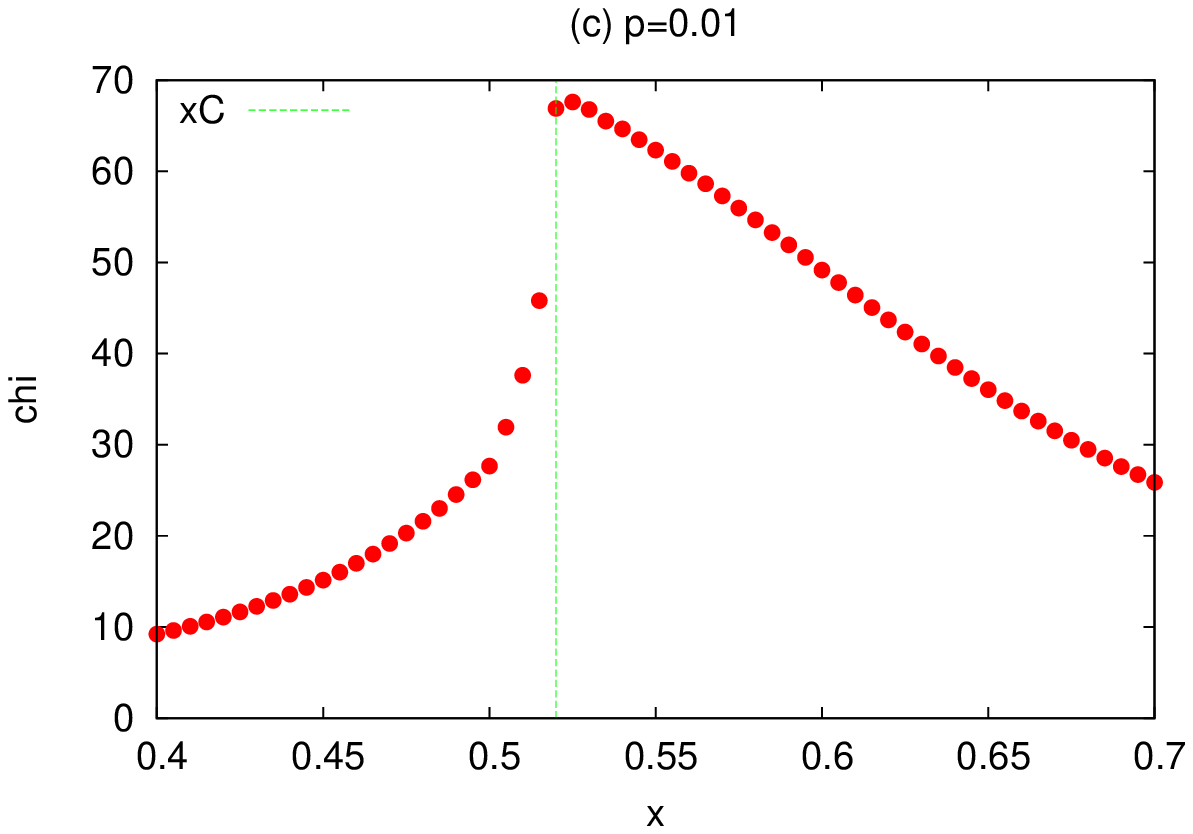}
\includegraphics[width=0.47\textwidth]{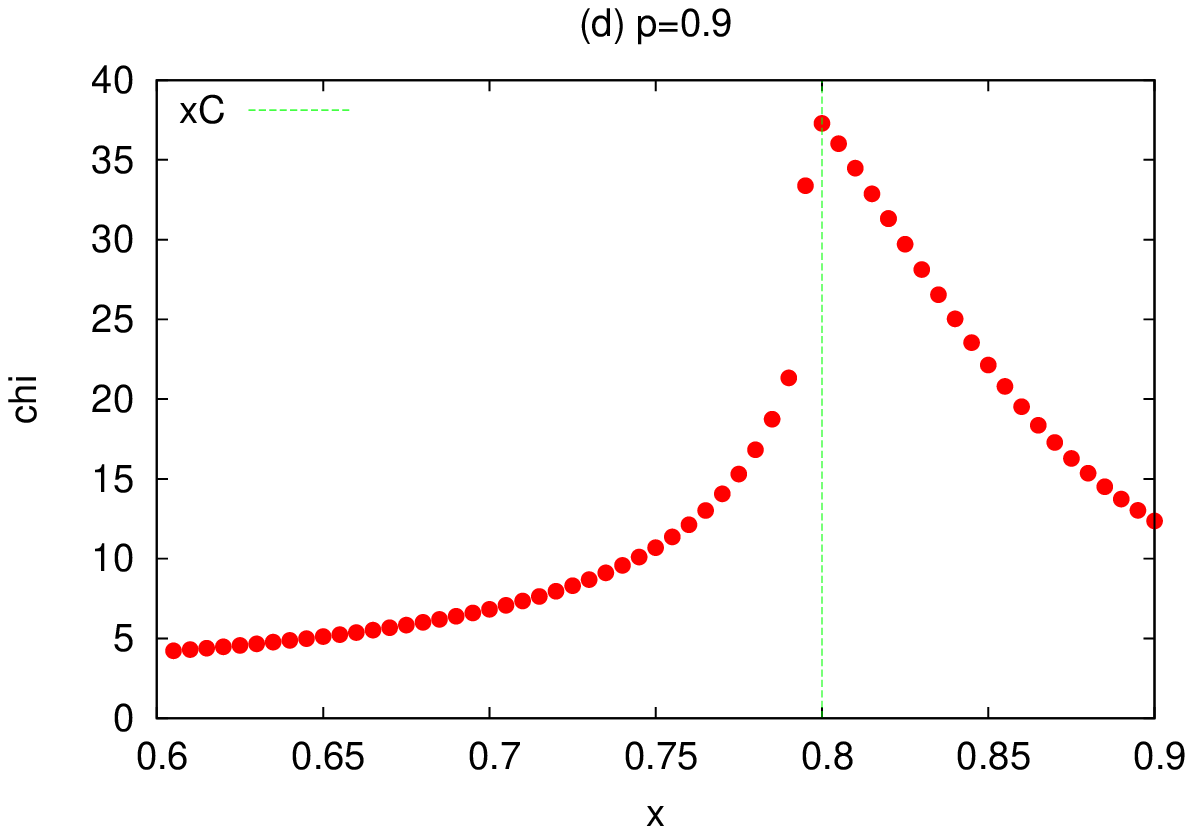}
\end{center}
\caption{\label{U4chi} The dependence of the Binder cumulant $U_4$ (a, b) and pedestrians' susceptibility for changing their strategy $\chi$ (c, d) on the model control parameter $x$.
The values of the susceptibility $\chi$ are obtained for $N=512^2$.
The vertical lines correspond to critical point position $x_C$.} 
\end{figure*}

Also the pedestrians' susceptibility for changing the strategy $\chi$ dependence on parameter $x$ may be used for critical point estimation.
For finite but large enough  system sizes $N$ the $\chi(x)$ dependence have maximum near $x_C$.
This maximum positions for $p=0.01$ and $p=0.9$ are marked by vertical lines in Figs. \ref{U4chi} (c, d).

The intersection points of the cumulants $U_4$ for different system sizes $N$ usually depend only rather weakly on those sizes, providing a convenient estimate for the value of the critical point $x_C$.
This intersection appears for $x_C\approx 0.52$ and for $x_C\approx 0.80$ for $p=0.01$ and $p=0.9$, respectively.
These intersection points coincide very nicely with points of vanishing order parameters $\langle m^k \rangle$ ($k=1,2$).

\section{Conclusions}

In this paper the influence of the long-range interactions on strategy selection was investigated. The critical point value $x_C$ increases monotonically with number of rewired links.
Critical point values $x_C$ indicated by $U_4(x;L)$ and $\chi(x)$ dependencies on parameter $x$ (Fig.~\ref{U4chi}) coincide nicely with $x_C$ evaluated from $\langle m\rangle(x)$ and $\langle m^2\rangle(x)$ dependencies (Fig.~\ref{mvsx}).
As we see, the athermal character of the model preserves the validity of the tools, commonly accepted in statistical mechanics.
Yet, it does not destroy typical system behaviours near the order--disorder critical point.

In our interpretation, the ordered phase is a model equivalent of a situation, where most of pedestrians accept the same strategy, selfish or cooperative. 
The result indicate, that a small amount of rewired bonds strongly supports the ordered phase.
This means in particular, that using mobile phones enhances the homogeneity of the strategy of the majority.
We note that a similar problem of interacting nodes in a network has been considered in \cite{jh1,jh2}, where spin-flip probabilities have been calculated within the Ising model.
There, the applied formulae rely on the well-known analogy with magnetic energy and temperature.
Our formulation and results allow to expect that most of these approaches can be reformulated within a more general, athermal frame.

\acknowledgments{The work was partially supported by the Polish Ministry of Science and Higher Education and its grants for Scientific Research and by the \href{http://www.plgrid.pl/en}{PL-Grid Infrastructure}.}


\end{document}